\documentclass[]{article}
\usepackage{amsmath,amssymb,amsfonts}
\usepackage{graphics,graphicx}
\usepackage[unicode,colorlinks,linkcolor=blue]{hyperref}
\usepackage{appendix}
\title{How to make the physical parameters small}
\author{Sergey G. Rubin \\
	National Research Nuclear University "MEPhI", \\ (Moscow Engineering Physics Institute),  \\
	N.I. Lobachevsky Institute of Mathematics and Mechanics,\\
	Kazan  Federal  University, \\
	Kremlevskaya  street  18,  420008  Kazan,  Russia\\
	sergeirubin@list.ru}

\begin{document}

\maketitle

\begin{abstract}
We study the evolution of the physical parameter values defined at the sub-planckian energies to values at low energies. The Wilson action is the basis of the research.
The presence of the compact extra dimensions has two consequences.
The positive point is that the integration over extra dimensions is a promising way to substantially reduce the parameters to be comparable with the observational values. On the other hand, the discreteness of the energy levels of compact extra dimensions complicates the analysis. This difficulty can be overcome with the truncated Green functions.
\end{abstract}

\section{Introduction}

It is assumed that the physics is formed at high energies \cite{Loeb:2006en,Ashoorioon:2013eia} - the widespread idea on the inflationary processes at the early Universe is the well-known example. Absence of the Quantum Gravity forces us working at the sub-planckian energies as the highest scale. 
The purpose of future research is to discover or guess the structure of theory at the sub-planckian scale - its dynamical variables, symmetries and the parameter values \cite{Brandenberger:2006vv,Tegmark:2005dy}. 

The evident difficulty is that we live at low energies, where the values of the observable physical parameters could strongly deviate from their initial values \cite{Chaichian:2000az}. Moreover, the same structure of the Lagrangian at high energies could be different from those at low energies due to the quantum corrections, for example.

One of the aims of the fundamental physics is to postulate a Lagrangian depending on primary parameters and fix them using their connection with observational values. This means that we have to calculate the parameter values at low energies starting from those at high energies \cite{Hertzberg:2015bta,Babic:2001vv,Dudas:2005gi}. The well known Wilson approach \cite{Burgess:2013ara} is accommodated for this purpose and this idea is used in this paper. 

The parameters at sub-planckian energies are usually supposed to be of the order of the Planck scale \cite{Ringeval_2019}, much larger than the observable values at the present time. At the same time, the variation of the parameter values with energy is proportional to coupling constants which are usually assumed to be small compared to the Planck scale. This means that there is no proper way to connect the parameters at both scales and hence new mechanisms should be involved which is the essence of the Hierarchy problem \cite{ArkaniHamed:1998rs,Guo:2018uxx}.  

The idea of extra dimensions is very popular now and is used in the solution to many problems \cite{ArkaniHamed:1998rs,1999PhRvL..83.3370R,Shifman:2009df}. It is shown here that the insertion of an extra space facilitates the connection of high energy Lagrangian structure and the low energy one. This is shown in a study done in our first article \cite{Rubin:2016ude}. At the same time, the extra space inclusion complicates the calculations, especially if the extra dimensions are compact. 

In the previous paper  \cite{Rubin:2016ude}, we studied the role of the compact extra dimensions on the parameter variation with energy. Starting from a moderate scale, we have shown that integration over additional coordinates can lead to a strong change in physical parameters from a 4-dimensional point of view. It is also possible to adjust them to the observable values with the help of the inhomogeneous extra dimensions \cite{Rubin:2015pqa}. 

In this paper, we continue discussion starting from the sub-planckian energy scale where the excitations of the extra space metric can not be neglected.
We start with the analysis of running constants at the high energies \cite{Marian:2019nhu} in the spirit of the Wilson action. The truncated Green function is used to neutralize the discreteness of the Kaluza-Klein spectra. As is shown below, the corrected parameters (the Lambda term is important exception) differ slightly from the initial ones if coupling constants are assumed to be small. This means that this mechanism is not very effective and can not be responsible for the parameter variation in many orders of the magnitude. It is also confirmed that integration over the extra-dimensional coordinates could be able to shift the parameter values to the observational range.

The analysis made in this paper is based on the toy model - scalar-tensor gravity \cite{Nojiri_2017,DeFelice:2010aj} with compact extra space. We start from the scalar field theory at high energies (small distances) and find the mass of the scalar field oscillations at lower energy scales. 

In the next Section, we check a statement on the smallness of the quantum corrections in the presence of the compact extra dimensions with its discrete spectrum. The Wilson approach is modernized because it is not clear how to extract an "infinitely small energy layer" in this case. 

\section{Parameters renormalization in presence of extra dimensions}

We use the Wilsonian approach to study the renormalization of the physical parameters in coarse of slowing down from high to small energies. Primary parameters are assumed to be fixed at high energies, $M_0$ in our case. There are several scales that influence dynamic of the parameters - D-dim Planck mass $m_D$, the electroweak scale $v$, the size of local extra space $r$ and the scale $M$. The relations between scales are assumed to be as follows: 
\begin{equation}\label{scales}
v\ll 1/r \ll M_0 < m_D
\end{equation}

In this Section, it will be shown that the perturbative procedure of the renormalization cannot help in a significant reduction of the physical parameters, the cosmological constant is the important exception. 

Here we study the energy dependence of the parameters in the spirit of the Wilson approach by defining physics at the highest energies. Descending to the electroweak scale $v$ or lower is the necessary step. This process is accompanied by a variation of physical parameters due to quantum fluctuations.
It appears to be not easy to shift energy to small values on the basis of pure gravity. Insertion of scalar field(s) improve the situation  \cite{Rubin:2016ude,Lyakhova:2018zsr}. 
 
Consider  the action of the scalar field and $f(R)$ gravity
\begin{eqnarray}\label{act}
&&S=\frac{m_D^{D-2}}{2}\int d^D z \sqrt{|g_D|}\left[f(R)+\partial_A \chi\partial^A \chi -\frac 12m^2\chi^2 - \lambda\chi^4\right], \\
&& f(R)=R+aR^2 + c,\quad m_D=1 \label{f}
\end{eqnarray}
acting in a D-dim space. The scalar field influence to a metric is neglected in this study.
The metric is chosen in the form
\begin{equation}\label{metric00}
ds^2 = g_D(z)_{AB}dz^Adz^B = dt^2 - e^{2Ht}(dr^2+r^2d\Omega^2_2) - e^{2\beta(y)}d\Omega^2_n.
\end{equation}
To facilitate the analysis,  we will assume $H\ll 1$ that allows us to use the Green functions in the Minkowski space. 

The generating functional for action \eqref{act} at the intermediate scale $M$	
\begin{eqnarray}\label{ZM}
&& Z_0^{M}=\int_0^{M} [D\chi]_{M} \exp\left(iS\right).
\end{eqnarray}
plays the central role in the Wilsonian approach. The extra space is supposed to be maximally symmetric, its radius $r$ and the corresponding energy scale $M_e$ are related as $M_e = 1/r$.
The functional \eqref{ZM} is the result of integrating out quick modes $M<m_D$ of the scalar field. The D-dimensional Planck mass $m_D$ is considered as the maximal energy scale in the rest of the article. 

The physical parameters at the scale $M_0\lesssim m_D$ are of the order of $m_D$. The reason for such choice is that high energy quantum corrections from the scales $M_0$ to the highest scale $m_D$ are assumed to have been taken into account in expression \eqref{ZM}.

Next two subsections are devoted to discussion on the energy shifting from high values to small ones with the subsequent parameter renormalization in the presence of the extra space.

Let us clarify the notion "scale of integration". There is no problem to think in terms of space scale $l$ or the energy scale $M$ with the relation $l=1/M$ in 4 dimensions \cite{Peskin:1995ev}. The energy scale is also expressed in terms of the euclidean wave vector $\bf{k_l}$, $M=k_l=|\bf{k_l}|.$
The presence of local extra dimensions complicates these relations due to the discreteness of their energy levels.
There are two kinds of the scalar field excitations in this case - one of them is represented by the waves along the four dimensions. The second set of the excitations in the compact extra dimensions is discrete. The correspondence between the energy of excitations in compact space and the space scale is $M=K_l/r,\, K_l=0,1,2...$ for 1-dim extra space \cite{1997PhR...283..303O}. 

Suppose now that we wish to integrate over "very thin" momentum layer $\delta k$ in the 4 large dimensions. Immediate question is - which of the energy states related to the extra dimensions should be taken into account also? If a chosen energy interval does not contain any discrete level $K_l$ we should forget about integration over the extra space modes. We should keep them in mind only in those cases when $k_l-\delta k_l <K_l/r <k_l$. Therefore the differential equations that describe the renormalization flow of parameters can not be applied in this case.

Let us start with the standard Wilsonian procedure and formally divide the field into quick $\chi_q(z)$ and slow $\chi_s(z)$ parts
\begin{equation}\label{quickslow}
\chi(z)=\chi_s(z)+\chi_q(z)
\end{equation}
The quick modes relate to an energy interval $(M-\delta M \div M)$ which is not obligatory infinitesimally small.

Substitution \eqref{quickslow} into \eqref{ZM} gives the generating functional in the form (the units $m_D=1$ are used in the rest of the paper)
\begin{eqnarray}\label{Z0M0}
&& Z_0^{M}  =Z_0^{M-\delta M}\cdot Z^M_{M-\delta M} \\
&&Z^M_{M-\delta M}=\int^{M}_{M-\delta M}D\chi_q \exp{\{\frac i2 \int d^Dz\sqrt{g_D}\left[\frac12 (\partial\chi_q)^2 -\frac12 m^2\chi_q^2 -\delta U(\chi_q,\chi_s)\right]\} },\nonumber \\
&&Z_0^{M-\delta M}=\int_{0}^{M}D\chi_s \exp{\{\frac i2\int d^Dz\sqrt{g_D}\left[\frac{1}{2}(\partial\chi_s)^2-U(\chi_s)
	\right]\}} \nonumber
\end{eqnarray}
where 
\begin{equation}\label{dU}
\delta U(\chi_q,\chi_s)=\lambda\chi_q^4+4\lambda\chi_q^3\chi_s  
+6\lambda\chi_q^2\chi_s^2
+ 4\lambda\chi_q\chi_s^3+ \lambda\chi_s^4.
\end{equation} 
Here we have taken into account orthogonality of $\chi_{s}$ and  $\chi_q$.

{The way to integrate out the field $\chi_q$ from \eqref{Z0M0}, provided that the coupling constant $\lambda$ is small, is well known (see for example textbook \cite{Peskin:1995ev})}. The result of integration over quick modes can be written in the Euclid space as \cite{Das, Peskin:1995ev}
\begin{eqnarray}\label{ZJ0}
&&Z^M_{M-\delta M} =\exp{\left[-\int d^Dz \sqrt{g_D}\delta U\left(\frac{\delta}{\delta J(z)},\chi_s(z)\right)\right]}
\cdot Z_J^{(E)}.
\end{eqnarray}
Here the generating functional in the  Euclid space
\begin{equation}\label{ZJ}
Z_J^{(E)}=\int_{M-\delta M}^M D\chi \exp{\left(-\int  d^Dz\sqrt{g_D}\left[\frac12 (\partial\chi)^2 +\frac12 m^2\chi^2 -\chi J\right]\right) }
\end{equation}
is a functional of an external current $J$. Integration over small deviations $\delta \chi_q = \chi_q-\chi_c$ around the classical part of the field gives 
\begin{eqnarray}\label{ZJ1}
&&Z_J^{(E)}=\exp\{-\frac12 \sum_{N\subset\cal{N}}\ln(\lambda_N+m^2)\}\cdot \\
&&\exp\{\frac12 \int d^Dz\sqrt{|g_D(z)|} d^Dz'\sqrt{|g_D(z')|}J(z)G'(z,z')J(z')\} \nonumber
\end{eqnarray}
The set $\cal{N}$ is a set of eigenstates that were integrated out. The complex index $N=(k,K)$ where $k$ is the wave vector in the 4-dim Euclidean space and $K$ marks eigenstates in the compact extra space and represents a discrete set. $G'(z,z')$ is the truncated Green function discussed in the Appendix. This Green function is cut both from UV and IR sides because the integration over finite energy interval is assumed.

Let us suppose that we have integrated out all discrete modes $K$ for which $K\geq K_l$. This means that we are disposed at the energy scale is $M=K_l/r$ with space scale $l=r/K_l$ and the 4-dim mode of the energy excitation
\begin{equation}\label{rel}
k_l=1/l=K_l/r.
\end{equation} 
Now, we have to choose next coarse-graining space scale $l+\Delta l$ and subsequent energy shift. We define it by the algebraic equation
\begin{equation}\label{Kl}
K_{l+\Delta l}=K_l-1 \rightarrow \Delta l=\frac{r}{K_l(K_l-1)}.
\end{equation}
The last expression follows from \eqref{rel}. The meaning of \eqref{Kl} is as follows: the selected integration layer $\Delta l$ relates to one and only one number $ K_l $ of energy state by definition. The size of the 4-dim energy layer is
\begin{equation}\label{kl}\Delta k_l\equiv k_{l}-k_{l+\Delta l} =1/r
\end{equation}
according to relations \eqref{rel} and \eqref{Kl}. Therefore, the contribution of one step of the coarsening $l\rightarrow l+\Delta l$ is as follows
\begin{eqnarray}\label{sum}
\Delta_l ... = \sum_{N\subset\cal{N}}...=\sum_{K=0}^{K_{l+\Delta l}}\sum_{k=k_l}^{k_{l+\Delta l}}... +
\sum_{k=0}^{k_l} (K_{l+\Delta l} =const) ...
\end{eqnarray}
The first term in the sum represents the contribution of all modes that are quick in the 4-dim space. 
The last term includes all 4-dim modes containing only one excitation $K_{l+\Delta l}$ of the extra dimensions. All these excitations contain wavelengths in the interval $(l\div l+\Delta l)$. 
Manipulations with the expressions similar to \eqref{sum} are discussed in Appendix A in light of the truncated Green functions calculations.

Let us fix scales for the following discussion and estimations. Let $M_{Pl}\simeq m_D=1$, the extra space size is $r=10^6$ and the initial energy scale is $M_0=10^{-2}$. Therefore, the energy layer $\Delta k_l=1/r$ is much smaller than $M_0$. In this case, summation can be approximated by integration with appropriate accuracy.

Now we will show that correction to the mass $m$ also contains the small parameter $\lambda$ and hence is small. To verify this, let us estimate quantum corrections produced by terms proportional to $\chi_s^2$.  The latter can be extracted from \eqref{dU}, \eqref{ZJ0}  and has the form
\begin{equation}\label{dU2}
\delta U_2 \equiv {6\lambda}\chi_q(z)^2 \chi_s(z)^2.
\end{equation}
Receipt \eqref{ZJ0} with $\delta U_2 $ instead of $\delta U$ leads to the quantum correction 
\begin{equation}\label{dU2s}\delta U_2(\chi_s)= \frac32 \lambda G'(z,z) \chi_s(z)^2\end{equation}
to the potential in the first multiplier $Z_0^M$ of expression \eqref{Z0M0}.
Therefore, the the correction to the scalar field mass equals
\begin{equation}\label{Dm2}
\Delta_l m^2= 3 \lambda G'(z,z)
\end{equation}
and is small if the coupling constant is small.
The truncated Green function has the form (see Appendix A)
\begin{eqnarray}\label{Green1}
&& G'(z,z)=\frac{1}{8\pi^3 r} \Big[\frac{k_l^3}{m_K}\arctan\Big(\frac{k_l}{m_K}\Big)\\
&&+m_K^2\ln(m_K)+\frac12 k_l^2-\frac12 m_K^2\ln(k_l^2+m_K^2)\Big] \nonumber
\end{eqnarray}
Here $m_K^2=k_l^2 + m^2$ and $k_l=M$ is the running energy scale. Substituting the Green function \eqref{Green1} into the expression \eqref{dU2s} we obtain the quantum correction to the scalar field mass

The summation over layers gives
\begin{equation}\label{Dm2tot}
\Delta m^2 =3\lambda  \sum_{k_l=1/r}^{M_0} G'(z,z) \simeq 3\lambda r \int_{1/r}^{M_0}dk_l G'(z,z) .
\end{equation}

According to \eqref{Green1}, the large parameter $r$ is canceled and the quantum correction appears to be proportional to the coupling constant $\lambda \ll 1$. All other values like $M, m$ are smaller than unity ($m_D=1$) and does not spoil this conclusion. 

The expression \eqref{Dm2} represents the mass difference between observations at two energy scales - $M$ and $M-1/r$. Expression \eqref{Dm2tot} is obtained by the integration over all layers in the energy interval $M_0>M>1/r$. The lower boundary is chosen because there are no KK excitations below $1/r$. Renormalization of the KK masses due to the quantum effects were studied in \cite{Bauman_2012}.

Calculation of the change in the energy density of the ground state (first line in expression \eqref{ZJ1}) during the one-step transition from the energy scale $M\equiv k_l$ to the scale $k_{l+\Delta l}=k_l + 1/r$ gives the following result
\begin{eqnarray}\label{groundstate}
&&\frac{\Delta \epsilon_l}{V_5} \simeq \frac{1}{16\pi^3r }  \Big[ - k_l^4\ln(k_l^2+m_K^2)+2k_l^4-2m_Kk_l^3\arctan\Big(\frac{k_l}{m_K}\Big) \\
&& -\frac12 m_K^4\ln m_K + \frac14(m_K^4- k_l^4)\ln(m_K^2+k_l^2)-\frac14 k_l^2m_K^2+\frac18 k_l^4\Big]. \nonumber 
\end{eqnarray}
Here $V_5=2\pi r V_4$ is the 5-dimensional volume. The result was obtained by combining expressions \eqref{CC4} and \eqref{int} in Appendix B.

Let us estimate the value of the total energy density shift provided that the energy scale is reduced from the initial scale $M_0$ to the scale $1/r$ where the fluctuations in the extra dimensions are suppressed. To this end, we use expression \eqref{groundstate} to obtain
\begin{equation}\label{Drhotot}
\frac{\Delta \epsilon}{V_5} \simeq \sum_l\frac{\Delta \epsilon_l}{V_5} \sim \frac 1r\sum_{k_l=1/r}^{M_0} m^4 \ln m \sim \int_{1/r}^{M_0}  dk_l m^4 \ln m\sim M_0m^4\ln m .
\end{equation}
inequalities $k_l\leq M_0\leq m\leq m_D=1$ were used here.
The variation of mass $\Delta m$ was neglected due to its smallness proved above.
The energy density $M_0$ can not be larger than the Planck one $m_D=1$, hence the correction to the ground state energy density is small as compared to the initial value $\epsilon_0 \sim m_D=1$:
$$\frac{\Delta \epsilon}{V_5}\simeq M_0m^4 \leq m_D^5.$$
Evidently, such renormalization can not make the $\Lambda$-term be equal to the observational value which is almost zero.

\section{Moderate energies. Reduction to 4 dimensions}

It was shown in the previous section that the physical parameters vary slowly with the energy, at least for the usual assumption on the coupling constant $\lambda \ll 1$. Therefore, the shift of the parameter values from high energy to small energies is a serious problem. In this section, it will be shown that the extra dimensions supply us a more effective mechanism of parameters changing provided that the dimensionality of extra space $n\ge2$. We will assume $n=2$ in this section.

There are five energy scales - the D-dimensional Planck mass $m_D$, the initial energy scale $M_0$ where we intend to specify our Lagrangian, the scale $M_e$ which is related to a characteristic size $r=1/M_e$ of the compact extra dimensions, the electroweak scale $v$ where the parameter values are known and the running scale $M$. 

Excitations discussed in the previous section act in each of $D$ dimensions if $M_e <M$. The picture is simplified when the energy scale shifts below $M_e$ where the excitations of the extra space are suppressed and hence the extra metric is static. It is reasonable to integrate over the extra coordinates in this case. This strongly influences the values of the physical parameters as is discussed in this section.


Consider the scalar field action taken from \eqref{act}
\begin{eqnarray} \label{Ssc} 
&&S_{\chi}=\frac{1}{2}\int d^4  xd^n y\sqrt{|g_4 g_n|} [ \partial_{A}\chi(z)g^{AB}\partial_{B}\chi(z) \\
&& - \epsilon m^2 \chi(z)^2 - \lambda \chi(z)^4 ] ,\nonumber
\end{eqnarray}
where $\epsilon=\pm 1$. As was shown in \cite{Rubin:2015pqa,Gani:2014lka,2017JCAP...10..001B}, there exist static solution $\chi=\chi_{cl}(y)$ to the classical equation
\begin{equation}\label{claseq}
\square_D \chi +U'(\chi)=0
\end{equation}
which are homogeneous in our 4-dim space. Let us decompose the field around its classical part
\begin{equation}\label{xY}
\chi(x,y) =\chi_{cl}(y)+\delta\chi;\quad \delta\chi\equiv\sum_{k=1}^{\infty}\chi_k (x) Y_k(y)\ll \chi_{cl}(y),
\end{equation}
where $Y_k(y)$ are the orthonormal  eigenfunctions of the d'Alembert operator acting in the $n$-dim extra space
\begin{equation}\label{eqY1}
\square_n Y_k(y)=l_k Y_k(y).
\end{equation}

Below, we limit ourselves by only first term in the sum \eqref{xY} so that
\begin{equation}\label{Y0}
\delta\chi=\chi_0(x)Y_0(y), \quad \square_nY_0=0,\quad Y_0=\frac{1}{\sqrt{v_n}}.
\end{equation}
Here $v_n$ is the volume of the compact extra space. This mode is distributed uniformly in the extra dimensions. After substitution \eqref{xY} and \eqref{Y0} into expression \eqref{Ssc} we get the following form of the effective 4-dim action for the scalar field $\chi_0(x)$
\begin{eqnarray}
S_{\chi}=\frac{1}{2}\int d^4  x \sqrt{|g_4|}[ \partial_{\mu}\chi_0(x)g^{\mu\nu}\partial_{\nu}\chi_0(x) 
- m_{eff}^2\chi_0 (x)^2 -2\Lambda_{\chi} +o(\chi_0^2) ]
\end{eqnarray}

where
\begin{eqnarray}
&&m_{eff}^2=\epsilon m^2 +\frac{6\lambda}{v_n}\int d^n y \sqrt{|g_n(y)|} \chi_{cl}(y)^2 , \label{m2}\\
&&\Lambda_{\chi}= \frac12\int d^n y \sqrt{|g_n(y)|} \left[-\partial_a\chi_{cl} (y)g_n^{ab}(y)\partial_b  \chi_{cl}(y) + \epsilon m^2\chi_{cl}^2 + \lambda \chi_{cl}^4 \right] \label{La}
\end{eqnarray} 
Our aim is to reduce the effective mass $m_{eff}$ in many orders of the magnitude provided that the initial mass $m\sim m_D =1$. One could imagine that equation
 $$m_{eff}^2=0$$ 
holds for a specific solution  $\chi_{cl}(y)=\tilde{\chi}_{cl}(y)$ to equation \eqref{claseq} provided that $\epsilon =-1$. Indeed, numerical simulation indicates that an appropriate function $\tilde{\chi}_{cl}(y)$ can be found for a wide range of the coupling constant $\lambda$ and the extra space volume $v_n$.  This means that we are able to substantially reduce the mass of the scalar field from the Planckian scale $\sim m_D\sim M_{Planck}$ to the electroweak one. Moreover, its value at low energies could have any sign. The Higgs-like potential with two minima is realized if the sign of the mass term remains negative.

The cosmological constant $\Lambda_{\chi}$ is also a function of the classical field distribution $\chi_{cl}(y)$ in the extra dimensions. As was shown in \cite{Rubin:2016ude}, one can choose a function $\bar{\chi}_{cl}(y)$ that gives zero value of the cosmological constant. Notice that $\tilde{\chi}_{cl}(y)\neq \bar{\chi}_{cl}(y)$ so that a more complicated extra metric should be used in future to simultaneously correct the mass and the Lambda term.

\section{Conclusion}
In this paper, we use the Wilsonian approach to study the evolution of the physical parameter values defined at the sub-planckian energies to values at low energies. It is shown that all parameters are varied slowly as compared to their initial values which are of the order of the Planck scale. This means that we need an alternative mechanism that is responsible for a significant change in the parameter values so that they coincide with the observational ones. We study the role of the local extra dimensions in this relation.

The inclusion of the compact extra space has two consequences. The positive point is that the integration over extra dimensions is a promising way to substantially decrease the parameter values to be comparable with the observable values. At the same time, the integration over small energy layers which is intrinsic to the Wilsonian approach appears to be problematic due to the discreteness of the Kaluza-Klein energy levels. In this regard, the truncated Green functions were used for analysis.

It is also shown that for any mass of scalar field at low energies, there exists an inhomogeneous extra metric that relates this mass value to its value at the Planck energy scale.



The integration over of the extra dimensions looks promising tool for a substantial shift of the parameter values to small ones at the low energy. A more general study based on the Green functions in the de Sitter metric \cite{Akhmedov:2019esv} looks promising.

\section{Acknowledgment}
The work was supported by the Ministry of Education and Science of the Russian Federation, MEPhI Academic Excellence Project (contract N~02.a03.21.0005, 27.08.2013) and according to the Russian Government Program of Competitive Growth of Kazan Federal University.


\appendix
\renewcommand{\thesubsection}{\Alph{subsection}}
\renewcommand{\theequation}{A\thesection.\arabic{equation}}
\section*{Appendix}

\subsection{Truncated Green function in the presence of the compact extra dimensions}


Here we consider the integration of generating functional \eqref{ZJ} over a finite set $\cal{N}$ of modes of excitation in the space with metric $M_4\times M_n$. The case $n=1$ is considered as a specific example. The excitations are described by the eigen-functions $Y_N(x,y)$ of the D'Alembertian operator
\begin{equation}\label{YN}
\square_{D}Y_N(z)=\lambda_N Y_N(z).
\end{equation}
The integration over a finite set of modes $\cal{N}$ assumes the decomposition
\begin{equation}\label{phis}
\chi(z)=\sum_{N\subset\cal{N}}\chi_{N}Y_N(z).
\end{equation}
Following the standard procedure for the calculation of generating functional \eqref{ZJ}, consider the solution 
\begin{equation}\label{solu}
\chi_c(z)=\sum_{N\subset\cal{N}}\frac{\int d^Dz'\sqrt{|g_D|} Y_N(z')J(z')Y^*_N(z)}{\lambda_N +m^2}
\end{equation}
to the equation
\begin{equation}\label{class}
\square_{D}\chi_c(z)+m^2\chi_c(z)=J(z)
\end{equation}
Comparison of expression \eqref{solu} with the definition of the Green function
\begin{equation}\label{Green2}
\chi_c(z)=\int d^Dz'\sqrt{|g_D|}G'(z,z')J(z')
\end{equation}
gives the relation between the truncated Green function $G'$ and the finite number of the eigen-functions
\begin{equation} G'(z,z')\equiv
\sum_{N\subset\cal{N}}\frac{ Y_N(z')Y^*_N(z)}{\lambda_N +m^2}
\end{equation}
If $\cal{N}$ is full orthonormal set then $G'$ represents the well-known form of the Green function. 

Let us refine the formulas written above and apply them to our needs.
The metric \eqref{metric00} leads to the decomposition
\begin{equation}\label{square}
\square_{D}=\square_{4}(x)+\square_{n}(y)
\end{equation}
which simplifies the expressions for the eigen-functions and eigen-values:
\begin{equation}\label{YN1}
Y_N(x,y)=Y_k(x)\cdot Y_K(y)
\end{equation}
and
\begin{equation}\label{lN}
\lambda_N=\lambda_k +\lambda_K
\end{equation}
The form of excitations in the 4-dim Euclidean space is known
\begin{equation}\label{lk}
\lambda_k = k^2,\quad Y_k(x)=\frac{1}{\sqrt{V_4}} e^{i(kx)}
\end{equation}
as well as in the 1-dim extra space
\begin{equation}\label{lK}
\lambda_K = K^2/r^2, \quad  Y_K(y)=\frac{1}{\sqrt{2\pi r}} e^{iKy/r}.
\end{equation}

Acting as in \eqref{sum}, the Green function  \eqref{Green2} acquires the form
\begin{eqnarray}\label{Green3}
&&G'(z,z) = \frac{1}{2\pi r V_4}\sum_{N\subset\cal{N}}(\lambda_N+m^2)^{-1}\\
&&=\frac{1}{2\pi r V_4}\left[\sum_{K=0}^{K_{l+\Delta l}}\sum_{k=k_l}^{k_{l+\Delta l}}(\lambda_k+\lambda_{K}+m^2)^{-1} 
 +\sum_{k=0}^{k_l} (\lambda_k+\lambda_{K_{l+\Delta l}}+m^2)^{-1} \right] \nonumber
\end{eqnarray}
Summation over 4-dimensional excitations can be replaced by integration in a standard way with the result
\begin{eqnarray}\label{Green4}
&&G'(z,z) = \frac{1}{4\pi^2\cdot2\pi r} \sum_{K=0}^{K_{l}-1}\int_{k=k_l}^{k_{l+\Delta l}}dk k^3 (k^2+K^2/r^2+m^2)^{-1} \nonumber \\
&&+\frac{1}{4\pi^2\cdot2\pi r} \int_{k=0}^{k_l}dk k^3  (k^2+(K_l -1)^2/r^2+m^2)^{-1}  
\end{eqnarray}

To move ahead, we have to fix scales for the numerical estimations here and in the main text. Let $m_D\simeq M_{Pl}=1$, the extra space size is $r=10^6$ and the primary energy scale is $M_0=10^{-2}$. Then, the number of extra space modes (or the Kaluza-Klein levels) is $K_l=M_0/r^{-1}=10^4$ and we may replace the sum over $K$ by the integral. Those part of the Green function that contains quick mode in 4-dim space (first line above) acquires the form
\begin{eqnarray}
&&G'_{4dim}\simeq \frac{1}{8\pi^3 r}\sum_{K=0}^{K_{l}-1}\Delta k_l k_l^3(k_l^2+K^2/r^2+m^2)^{-1} \nonumber \\
&&\simeq \frac{1}{8\pi^3}k_l^3\Delta k_l\int_0^{(K_l -1)/r} d\zeta (k_l^2+\zeta^2 +m^2)^{-1} \nonumber \\
&& \simeq  \frac{1}{8\pi^3}\frac{k_l^3}{r}\int_0^{k_l} d\zeta (k_l^2+\zeta^2 +m^2)^{-1} \nonumber \\
&& =\frac{1}{8\pi^3}\frac{k_l^3}{rm_K}\arctan\Big(\frac{k_l}{m_K}\Big)
\end{eqnarray}
Here  $m_K^2=k_l^2 + m^2$, $\Delta k_l = 1/r$ and $K_l - 1\sim K_l=rk_l$.

The second line in \eqref{Green4} is responsible for the only quick mode acting in the extra dimensions:
\begin{eqnarray}
&&G'_{extra}\simeq \frac{1}{8\pi^3 r} \int_{0}^{k_l}dk k^3  (k^2+m_K^2)^{-1} \nonumber \\
&&=\frac{1}{8\pi^3 r}\left[m_K^2\ln(m_K)+\frac12 k_l^2-\frac12 m_K^2\ln(k_l^2+m_K^2)\right].
\end{eqnarray}
and the truncated Green function at the energy level $k_l$ is as follows
\begin{equation}\label{Green5}
G'(z,z)=G'_{4dim}+G'_{extra}.
\end{equation}


\subsection{ Ground state energy in the presence of the compact extra dimensions}
\renewcommand{\theequation}{B\thesection.\arabic{equation}}
The shift of the ground state energy (the first exponent in \eqref{ZJ1}) is represented in this Appendix. The contribution of layers in the interval $\Delta k_l$ from \eqref{kl} is as follows

\begin{eqnarray}\label{CC}
&&\Delta\epsilon = -\frac12 \sum_{N\subset\cal{N}}\ln(\lambda_N+m^2)=\\
&&-\frac12 \sum_{K=0}^{K_{l+\Delta l}}\sum_{k=k_l}^{k_{l+\Delta l}}\ln(\lambda_k+\lambda_{K}+m^2) \nonumber \\
&&-\frac12\sum_{k=0}^{k_l} \ln(\lambda_k+\lambda_{K_{l+\Delta l}}+m^2)  \nonumber
\end{eqnarray}
The first term in the sum represents the contribution of the quick modes in the 4-dim space. 
The last term includes all modes in the 4-dim space - quick and slow - and quick excitation with the number $K_{l+\Delta l}$ acting in the extra dimensions. All these excitations are characterized by wavelength in the interval $(l\div l+\Delta l)$ in 4 dimensions or in the extra space. 
The sums over $k$ and $K$ can be converted into the integral in the standard form
$$\sum_k=\frac{V_4}{8\pi^2}\int dk k^3, \quad \sum_K =r\int dK$$
and we get the Euclidean version of the energy density
\begin{eqnarray}\label{CC2}
&&\frac{\Delta \epsilon_l}{V_5} = -\frac{1}{16\pi^3 r}\Bigg[ \sum_{K=0}^{K_{l+\Delta l}}\int_{k=k_l}^{k_{l+\Delta l}}dk k^3 \ln (k^2+\lambda_{K}+m^2)  \\
&&+ \int_{k=0}^{k_l}dk k^3  \ln(k^2+\lambda_{K_{l+\Delta l}}+m^2)\Bigg]. \nonumber
\end{eqnarray}
Here $\lambda_k = k^2$ and $V_5=2\pi r V_4$ is the 5-dimensional volume. $V_4$ is 4-dim volume in the Euclidean space.
The interval of the energy layer $\Delta k_l$ is assumed to be small and we neglect variation of the physical parameters within this interval. Then \eqref{CC2} is
\begin{eqnarray}\label{CC3}
&&\frac{\Delta \epsilon_l}{V_5} \simeq -\frac{1}{16\pi^3 r} k_l^3 \Delta k_l\sum_{K=0}^{K_{l+\Delta l}} \ln(k_l^2+\lambda_{K}+m^2)+ \\
&& \frac{1}{16\pi^3 r} [-\frac12 m_K^4\ln m_K - \frac14( k_l^4-m_K^4)\ln(m_K^2+k_l^2)-\frac14 k_l^2m_K^2+\frac18 k_l^4] \nonumber
\end{eqnarray}
where $m_K^2 = \lambda_{K_{l+\Delta l}} +m^2$.

Remind that we are integrating over the wavelength interval $(l\div \Delta l)$. 
Let us express everything in terms of the left edge of the interval $l$ or the wave number $k_l=1/l$.
The wavelength of the scalar field excitation in extra space is $l=r/K_l$. Hence $K_l=rk_l$. 

The right edge of the interval is defined as follows - $K_{l+\Delta l}=K_l-1$ that leads to the interval $\Delta k_l = 1/r$ and $m_K^2={(K_l-1)^2}/{r^2}+m^2\simeq {K_l^2}/{r^2}=k_l^2$ ($m\ll 1$)

\begin{eqnarray}\label{CC4}
&&\frac{\Delta \epsilon_l}{V_5} \simeq -\frac{1}{16\pi^3 }  \frac{k_l^3}{r^2} \sum_{K=0}^{K_l-1} \ln(K^2/r^2+k_l^2+m^2)+ \\
&& \frac{1}{16\pi^3 r} \Big[-\frac12 m_K^4\ln m_K + \frac14( k_l^4-m_K^4)\ln(m_K^2+k_l^2)+\frac14 k_l^2m_K^2-\frac18 k_l^4-\frac14 m_K^4\Big]. \nonumber
\end{eqnarray}

According to the discussion made in Appendix A, the sum in \eqref{CC4} can be approximated by the integral
\begin{eqnarray}\label{int}
&& \sum_{K=0}^{K_l-1} \ln(K^2/r^2+k_l^2)\simeq r\int_0^{K_l/r} d\zeta \ln (\zeta^2+k_l^2+m^2)  \nonumber \\ 
&& = k_l\ln(k_l^2+m_K^2)-2k_l+2m_K\arctan (k_l/m_K).
\end{eqnarray}
Finally, \eqref{CC4} is transformed in the following form
\begin{equation}\label{CC6}
\frac{\Delta \epsilon_l}{V_5} \simeq \frac{k_l^4}{16\pi^3 r}(\frac{15}{8}-\ln 2 - \frac{\pi}{2} - \frac52 \ln k_k).
\end{equation}


\end{document}